# CONTROL OF COMPLEX SYSTEMS USING BAYESIAN NETWORKS AND GENETIC ALGORITHM

**Tshilidzi Marwala**

*Centre for Systems and Control Engineering*
*School of Electrical and Information Engineering*
*University of the Witwatersrand*
*Private Bag 3, Wits, 2050, South Africa*

Abstract: A method based on Bayesian neural networks and genetic algorithm is proposed to control the fermentation process. The relationship between input and output variables is modelled using Bayesian neural network that is trained using hybrid Monte Carlo method. A feedback loop based on genetic algorithm is used to change input variables so that the output variables are as close to the desired target as possible without the loss of confidence level on the prediction that the neural network gives. The proposed procedure is found to reduce the distance between the desired target and measured outputs significantly.

## 1. INTRODUCTION

The control of engineering systems, such as bioprocesses, has been the subject of research for some time. A literature review on the subject can be found in Schurgel (2001). This paper reviews recent development of bioprocess engineering including monitoring of product formation processes. It also reviews advanced control of indirectly evaluated process variables by means of state space estimation using structured and hybrid models, expert systems and pattern recognition for process optimization. Control of engineering systems has been conducted in several areas, such as aerospace engineering, where it was applied to actively control pressure oscillations in combustion chambers (Blonbou, *et. al.,* 2000). Genetic algorithms and fuzzy logic have been successfully used to control load frequency in PI controllers (Chang, *et. al.,* 1998). Plant growth has been optimally controlled using neural networks and genetic algorithms (Morimoto and Hashimoto, 1996) and fuzzy controller has been used for active management of queuing problem (Fengyuan, *et. al.*, 2002).

The control procedure adopted in this paper consists of two components of a feedback control system. The first component is the forward component that takes the inputs and computes the outputs. In many complex problems, approximation methods, e.g. neural networks or fuzzy logic, are used to achieve this goal. The second component is the feedback loop that is only activated if the predicted output is not sufficiently close to the desired target. This is an optimisation problem and any number of optimisation tools such as gradient-descent methods can be used (Pallaschke, 1997). However, in many practical problems, which are generally complex, it is sometimes impossible to calculate the gradients required when gradient based optimisation methods are used. Furthermore, the gradient based methods are more susceptible to local optimum solutions rather than global optimum solutions. As a result of these limitations, over the past years evolutionary techniques such as genetic algorithms have become popular. This is due to the fact that they do not require gradients and are able to identify globally optimum solutions (Michalewicz and Dasgupta, 1997).

Thus far most control procedures that use neural networks use networks that are trained by the maximum-likelihood method (Bishop, 1995). Maximum-likelihood method for training neural networks is conducted by minimising the distance between the network training target and the neural network prediction. However, this procedure is only effective if the networks are trained in the conventional approach of training, validation and testing, which is not ideal for on-line control problems. Furthermore, the maximum-likelihood method does not give confidence levels on the predictions they give. As a result, the optimised input parameters do not necessarily fall within the learned input space and consequently, neural networks are not confident of the outputs they give. In this paper, an alternative neural network method, i.e. Bayesian neural networks are used to predict the output given the input data and be implemented in the context of control systems. The output predicted by neural networks also has confidence levels due to the Bayesian formulation. If the output is not sufficiently close to the target output, genetic algorithm is activated to sample the combination of input parameters that ensures that the predicted output is as close to the desired target output as possible. This is done such that the resulting predicted outputs retain high confidence levels.

The framework, proposed in this paper, is tested to optimally control the fermentation problem, which is a highly complex process. The reason why fermentation is chosen is because of its practical importance in areas such as pharmaceutical and food industries, which are vital for human life.

This paper makes the following contributions to the scientific literature: (1) contribute to control literature by tackling the control of highly complex scenario with multiple variables that involve biological organisms; (2) apply Bayesian statistics to ensure that the control algorithm is confident of the optimal solution it gives; (3) and apply genetic algorithm and Bayesian neural networks for the control of highly complex systems.

## 2. CONTROL FRAMEWORK

The control framework that is implemented in this paper is shown in Figure 1. As mentioned in the introduction, the first component of the framework is a feed-forward neural network, which takes input vector x given network weights w and predict output vector y as follows:

$$y = F(x, w) \tag{1}$$

The network weights in equation 1 are obtained through the learning process, which is explained in the next section. It must be borne in mind that the network weights, in this paper, form a probability distribution because we are employing Bayesian statistics to train the networks. As a result, the output vector also has a probability distribution from which confidence levels can be drawn.

The second component of the control loop, shown in Figure 1, is genetic algorithm optimiser. Its function is to identify input parameters that minimise the distance between predicted and desired target output vectors while ensuring that confidence levels on the prediction remain high. The objective function that is used to achieve this goal is:

$$\text{error} = \sum (y - t_d)^2 + (1 - CL) \tag{2}$$

In equation 2, y is the neural network output vector, $t_d$ is the desired target vector and CL is the confidence level and it ranges linearly from 0 for no confidence to 1 for full confidence. The CL is calculated from the average standard deviations of all the elements of the normalised output vector. The average standard deviation of 0 gives CL of 1 while an infinitely high standard deviation gives the CL of a 0. The design variables, in equation 2, are the input parameters to the neural networks. The second term, in equation 2, ensures that the identified input parameters that minimises the error fall within the subset of the information that has been learned before. This is crucial because neural networks only operate within the framework of the information they have learned before. The exclusion of CL tends to give the input parameters that when forward-propagated into the neural networks, they give inaccurate results. The details on solving equation 2 using genetic algorithm are explained later in the paper.

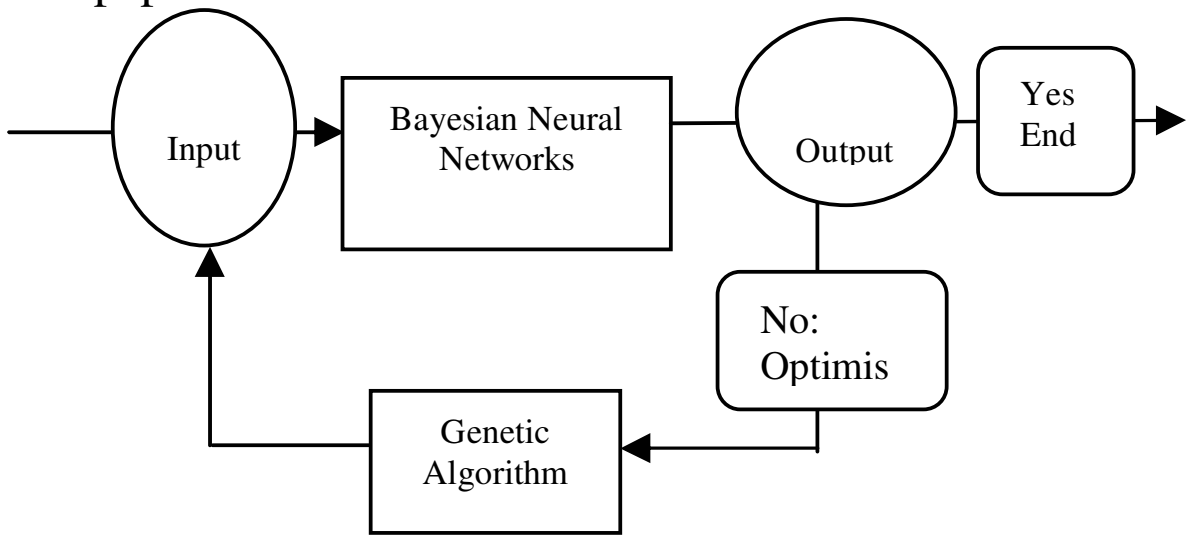

Fig. 1. Feedback control loop that uses Bayesian neural networks and genetic algorithm

## 3. NEURAL NETWORKS

As explained in the previous section, the first component of the control mechanism adopted in this paper is neural networks. Neural networks are tools that make probabilistic assumptions about data. Learning algorithms are methods for finding parameter values that look probable in the light of the data. In this paper, neural network learning is used to approximate the functional mapping between the input vector x and output vector y. In this paper, the multi-layer-perceptron (MLP), with a hyperbolic tangent basis function in the hidden units and linear basis functions in the output units, is used (Bishop, 1995). A schematic illustration of the MLP is shown in Figure 2 and the relationship between the $k^{th}$ output y and x may be may be written as follows (Bishop, 1995):

$$y_k = f_{outer}\left(\sum_{j=1}^{M} w_{kj}^{(2)} f_{inner}\left(\sum_{i=1}^{d} w_{ji}^{(1)} x_i + w_{j0}^{(1)}\right) + w_{k0}^{(2)}\right) \tag{3}$$

Here, M is the number of hidden units, d is the number of input units, $w_{ji}^{(1)}$ and $w_{ji}^{(2)}$ indicate weights in the first and second layer, respectively, going from input i to hidden unit j while $w_{j0}^{(1)}$ indicates the bias for the hidden unit j.

Training the network essentially means estimating the weight vector w that ensures that the output vector y is as close to the training target vector as possible. In this paper, Bayesian technique is applied to estimate the weight vector and this method is able to handle the lack of adequate amount of training data.

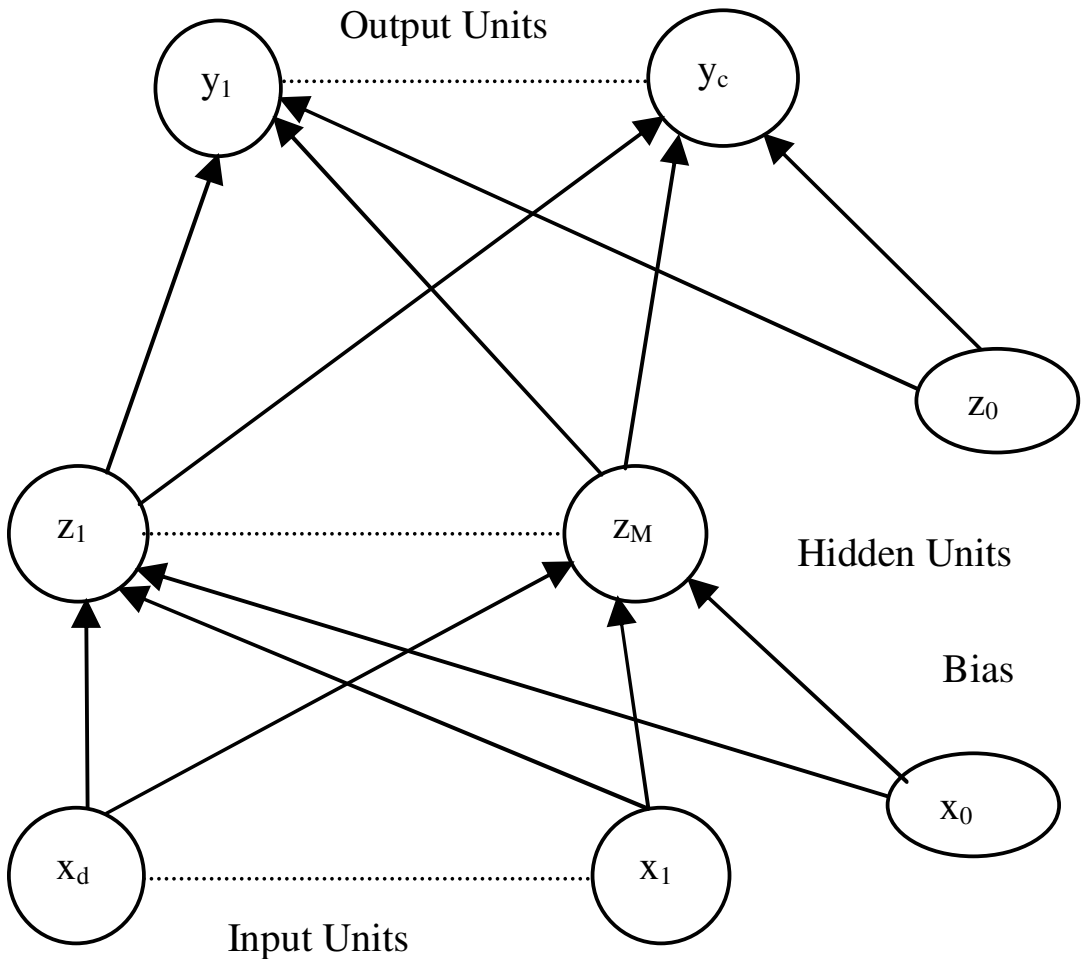

Fig. 2. Feed-forward network having two layers of adaptive weights.

The problem of identifying the weights ($w_i$) is posed in Bayesian form as follows (Neal, 1993):

$$P(w \mid D) = \frac{P(D \mid w)P(w)}{P(D)} \qquad (4)$$

where P(w) is the probability distribution function of the weight-space in the absence of any data, also known as the prior probability distribution function, and $D \equiv (y_1,\ldots,y_N)$ is a matrix containing the output data. The quantity P(w|D) is the posterior probability distribution after the data have been seen and P(D|w) is the likelihood probability distribution function while P(D) is the normalisation factor. Following the rules of probability theory, the distribution of output vector y may be written in the following form:

$$p(y \mid D) = \int p(y \mid w)p(w \mid D)dw \qquad (5)$$

For special cases, the distribution, in equation 5, may be calculated directly, however, for many practical problems it is estimated using Monte Carlo methods (Neal, 1993; Takaishi, 2002). The integral in equation 5 may, thus, be approximated as follows (Neal, 1993):

$$\tilde{y} \cong \frac{1}{L}\sum_{i=I}^{Q+L-1} F(w_i) \qquad (6)$$

Here $\tilde{y}$ is the estimated output, F is the neural network model that gives the output whenever the input is given, Q is the number of initial states that are discarded in the hope of reaching a stationary distribution represented by equation 5 and L is the number of retained samples. In this paper, the hybrid Monte Carlo method, which has been used quite extensively to solve complex engineering problems (Öcten, 2002), is used to estimate equation 5 through equation 6. The details of this technique, which are fairly abstract, are beyond the scope of this paper and can be obtained in (Neal, 1993). This technique is a form of a Markov chain with transition between states achieved by alternating the 'stochastic' and 'dynamic moves'. The 'stochastic' moves allow the algorithm to explore states with different total energy. The 'dynamics' moves are achieved by using Hamiltonian dynamics (Neal, 1993) and allowing the algorithm to explore states with the total energy approximately constant. This is achieved by following these steps: (1) Choose the step size ($\Delta w$) and the number of steps (L) in the trajectory; (2) From the initial weight vector ($w_{initial}$), take¸ L steps each of size $\Delta w$, in the weight space in the direction that result with higher posterior probability leading to vector $w_{current}$ [this direction is obtained by determining the gradient of p(w|D)]; and (3) If the posterior probability of the current sample is higher than from the previous sample, then accept $w_{current}$. Otherwise, select a random number $\xi$ of uniform distribution in the range [0, 1]. If $\frac{p(w_{current} \mid D)}{p(w_{old} \mid D)} \succ \xi$ then $w_{current}$ is accepted, otherwise it is rejected. This process is called Metropolis *et. al.* algorithm (1953).

## 4. GENETIC ALGORITHM

In Figure 1 it is indicated that the other component of the control process proposed in this paper is genetic algorithm. Genetic algorithms were inspired by Darwin's theory of natural evolution. In this paper, this natural optimisation method is used to optimise the cost function shown in equation 2. The genetic algorithm implemented in this paper uses a population of binary-string chromosomes (Holland, 1975). Each of these strings is the discretised representation of a point in the search space and, therefore, has a fitness function given by the objective function. On generating a new population, three operators are performed: (1) crossover; (2) mutation; (3) and reproduction.

The crossover operator mixes genetic information in the population by cutting pairs of chromosomes at random points along their length and exchanging over the cut sections. This has a potential of joining successful operators together. Simple crossover technique (Goldberg, 1989) is used in this paper. For simple crossover, one crossover point is selected, binary string from beginning of chromosome to the crossover point is copied from one parent, and the rest is copied from the second parent. For example, when **1100**1011 undergoes simple crossover with 1101**1111** it becomes 11001111.

The mutation operator picks a binary digit of the chromosomes at random and inverts it. This has a potential of introducing to the population new information. In this paper, binary mutation is used (Goldberg, 1989). When binary mutation is used, a number written in binary form is chosen, and its value is inverted. For an example: 11001011 may become 11000011.

Reproduction takes successful chromosomes and reproduces them in accordance to their fitness functions. In this paper roulette reproduction method is used (Goldberg, 1989). Roulette method can be viewed as allocating pie-shaped slices on a roulette wheel to population members. Each slice is proportional to the member's fitness. Selection of a population member to be a parent can then be regarded as a spin of the wheel. The winning population member is the one in whose slice the roulette spinner ends up. Even though this selection method is random, each parent's chance of being selected is directly proportional to its fitness. The least fit members are therefore gradually driven out of the population.

## 5. CASE STUDY

To validate the procedure proposed in this paper, the proposed method is tested to control fermentation. Fermentation is a process by which sugar is transformed into alcohol using yeast as a catalyst. There are many ways in which fermentation is controlled and this includes pursuing the chemical route (Johansson and Hahn-Hägerdal, 2002). This can be done by understanding the chemistry and adding chemical additives to control fermentation. The disadvantage of this procedure is that nowadays there are health pressure groups that have made it their missions for there not to be any added chemicals to goods that are consumed by human beings. The control of the fermentation process has been studied by O'Connor et. al. (2002), who used fuzzy-logic to control the fermentation process, with an objective being to find interrelationships between input and output variables. Their work was limited in the following ways: (1) the input parameters were not as comprehensive because fuzzy logic cannot handle many input variables; (2) fuzzy logic is more of an approximation method that neural networks and therefore tends to be less accurate; (3) fuzzy control scheme that was implemented is not able to optimally control a complex process, such as fermentation; (4) the control scheme proposed was not practically implemented. The present study addresses all these four issues mentioned above.

The device constructed, in this paper, to control the fermentation process is called the Fermentation Management System (FMS) and is illustrated in Figure 3.

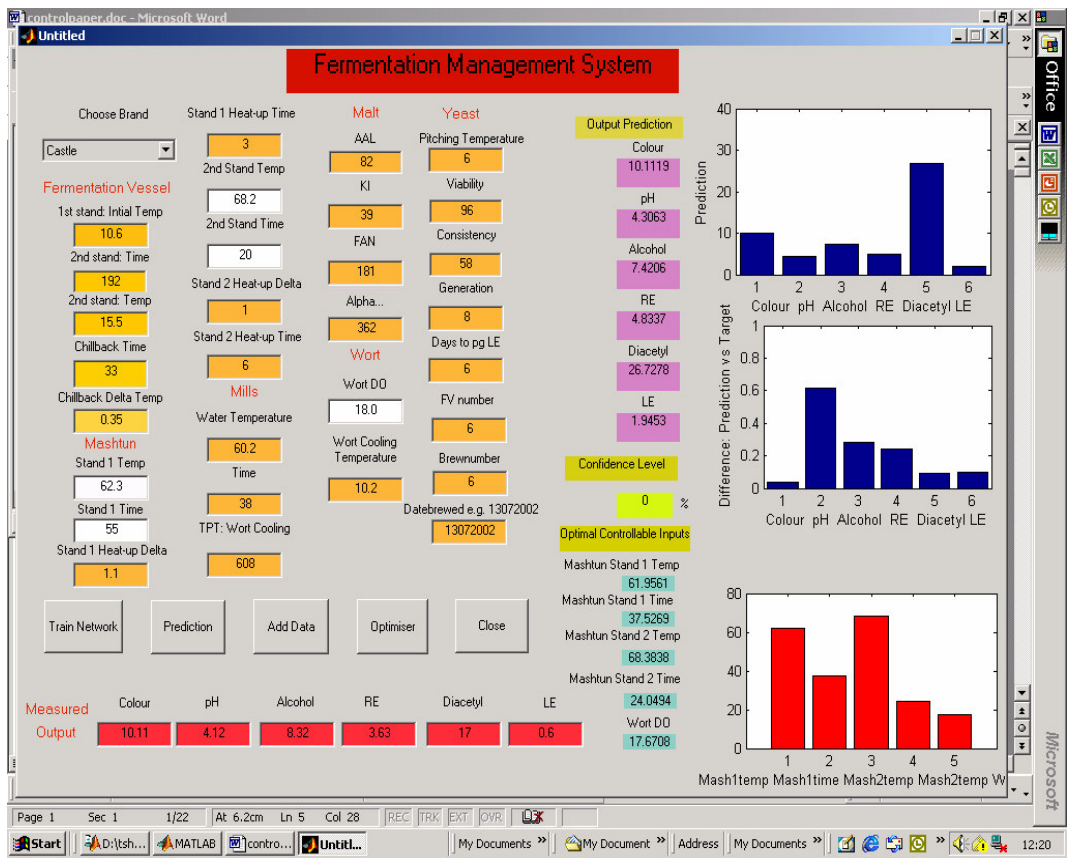

Fig. 3 Illustration of the fermentation management system implemented to control fermentation.

The FMS has the following components:

1. *Brand Chooser*: Here there are two beer brands to choose from.
2. *Input Parameters*: These are grouped into various types. The first group is the fermentation vessel and here the fermentation process occurs with yeast as a catalyst. The second group is the characteristics of the malt (which is obtained from barley through a process called malting, which is beyond the scope of this paper) from which fermentable sugars are obtained. Parameters in this group cannot be changed or modified and therefore do not form part of parameters that can be controlled. The third group is the mashtun where malt is heated through various temperatures for defined durations. There are some parameters in this group that can be changed for control purposes and these are stands temperatures and pressures. The fourth group is the mills where malt is meshed to prepare it for fermentation. In this paper, none of the parameters from the mills are conveniently controllable. The fifth group is the wort, which has the wort-dissolved-oxygen that can be controlled. Wort is filtered liquid sugar. The sixth group is the yeast which is essential for making of alcohol and is a single-cellular living organism with certain measurable properties. Input parameters that are controllable are two stands temperatures and their respective durations as well as the wort dissolved oxygen. These controllable parameters are highlighted in white colour in Figure 3.
3. *Predicted Output Parameters*: These are quality parameters that can be linked to the taste of the beer. These are beer colour, pH, alcohol; residual extract (RE), diacetyl and limit extract (LE).
4. *Graphical display of predicted outputs*: This makes it easy for the user to visually inspect the predicted output graphically as seen in Figure 3 with the y-axis named Prediction.
5. *Output numerical display*: This supports (4) and is indicated in Figure 3 (see Output Prediction).
6. *Difference between predicted outputs and targets*: This functionality allows the user to have some idea on how far the predicted output is from the desired target. The graph in the FMS can be used to manually control the fermentation process through trial-and-error. This is useful because brew-masters are reluctant to completely surrender the decision-making process to a computer program such as the FMS.
7. *Optimal controllable inputs*: These inputs, shown in white in Figure 3, can be controlled and are the mashtun 1 and 2 stands temperatures and durations as well as the amount of oxygen in the wort.
8. *Confidence levels (CL)*: This indicates the confidence the network has on a solution and allows a user to take or reject a recommendation of the FMS. Practical implementation of the FMS shows the CL of 80% as a cut-off point.
9. *Real measured outputs*: Measured outputs which are used to expand the training database.

The FMS infrastructure has the following pushdown buttons:

1. *Train Networks*: This activates a program that reads the database and trains the network and saves the characteristics of the trained networks.
2. *Prediction:* This takes the input data as well as trained networks' characteristics and predict the end of fermentation parameters.
3. *Add Data*: This takes measured output parameters and the corresponding input parameters and add them to the database.
4. *Optimiser*: This invokes the genetic optimiser to identify mash temperatures and the amount of dissolved oxygen that give predicted output that is as close to the desired target as possible
5. *Close*: This functionality closes the FMS

In many control problems the control procedure, such as the one proposed by Blonbou et. al. (2000) to control combustion, time is critical and a control algorithm has to be invoked within a fraction of a second. For these types of applications Bayesian networks are not suitable, due to the fact that they are computationally intensive relative to other types of neural networks. As a result, the advantage of confidence levels that, is offered by Bayesian networks is not exploited. However, for the present application, of optimal control of fermentation, time is not as critical because the fermentation process is a slow process that takes 10 days to complete. The window period in which the control process can be activated is one day.

In this paper, the architecture of the neural network constructed has 29 input parameters, 19 hidden units and six output units. The details of the input and output units are described at the beginning of this paper (see equation 3) and the activation units in the hidden layer is a hyperbolic tangent function while in the output layer is a linear function. The number of retained samples that form the posterior probability indicated in equation 4 is 500, while the number of discarded samples is 100. Some samples are discarded, as a matter of good practice, because they may not necessarily reflect the true posterior probability (see equation 4) due to the fact that the algorithm may not have reached sampling at the regions of the desired distribution. Bayesian neural network is used to estimate the predicted output, through equation 6, and the confidence level (CL) is obtained from the standard deviation of the distribution of the output (see equation 5). The standard deviation is normalised so that when standard deviation is 0 then the confidence level is 1. On training the neural networks 600 samples were used.

The genetic algorithm is constructed using the objective function in equation 2. The design variables are the five controllable parameters which are stands 1 and 2's temperatures and durations in the mashtun as well as the amount of dissolved oxygen in the wort. The size of the population of possible input parameters when implementing genetic algorithm is 40. The input parameters to be optimised are transformed from floating point to 16-bit format using Gray coding (Michalewicz and Dasgupta, 1997). The chromosomes (individuals in the population) represented by binary numbers are allowed to interact by using simple crossover procedure as mentioned above. The probability of crossover occurring is set to 0.6. This value was determined by trial and error. Each chromosome mutates at a probability of 0.0333. Again this value was obtained through trial and error. The chromosomes are transformed back to floating-point parameters. The objective function, given by equations 2, is then used to evaluate the fitness of each population member. The population members that are relatively fit are allowed to reproduce and the weak members are gradually eliminated using roulette wheel (Holland, 1975; Goldberg, 1989) procedure. When the fitness of the population has converged then the procedure is terminated. Otherwise, the process of crossover, mutation and reproduction is repeated. It was observed that convergence was generally achieved after 20 genetic algorithm generations.

The computational time taken to train the neural network on a Pentium 3 with 200MHz of RAM was 5 computer processing unit (CPU) minutes, while the time it took to determine the optimum input parameters using genetic optimiser was 6 CPU minutes. However, the time taken to predict the output was 3 CPU seconds while the time taken to add an additional sample to the database was 0.5 CPU second.

## 6. DISCUSSION

The Bayesian neural network prediction is shown in Figure 4. This is an average prediction as obtained by using equation 6. Figure 5 shows results of the prediction versus actual that was taken to asses the accuracy of the neural networks. These figures show that indeed Bayesian neural networks offer accurate prediction of the end of fermentation parameters.

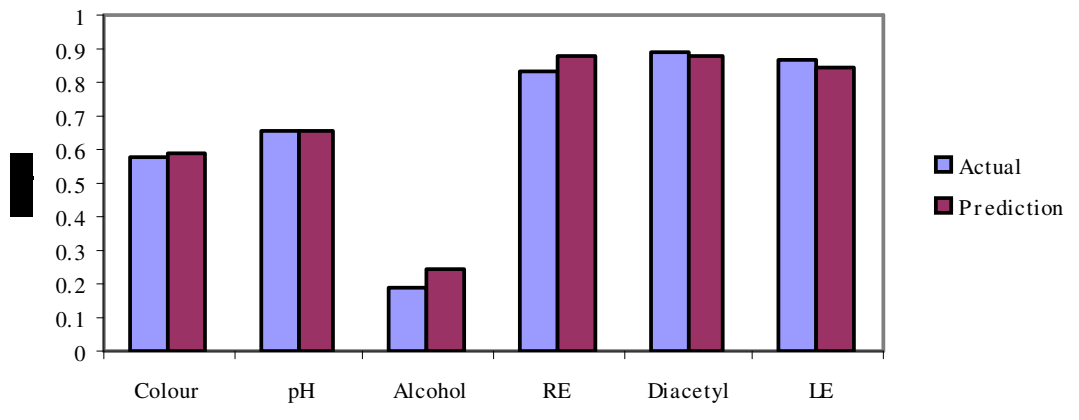

Fig. 4. Results of the output of the neural networks Key: RE: residual extract; LE: limit extract

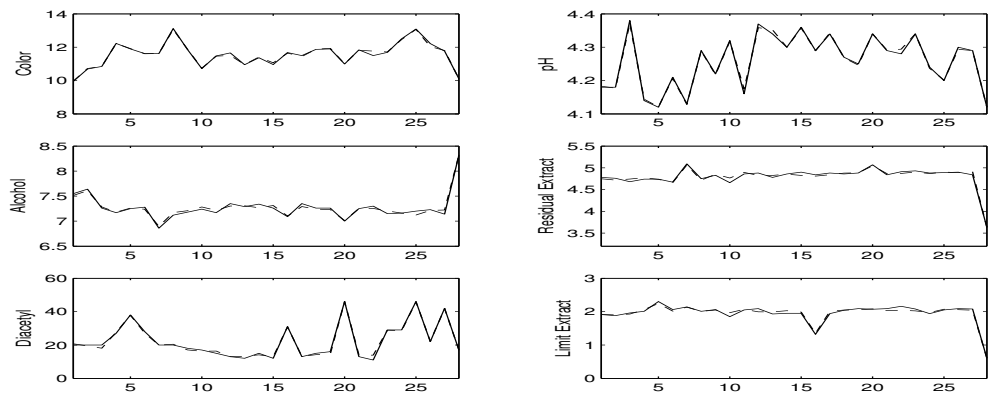

Fig. 5. A snapshot of 29 samples showing the output of the neural networks

The sample distribution of the output corresponding to colour is shown in Figure 6. From this figure the average output calculated through equation 6, is 0.35. From this figure several characteristics of the output may be derived including the standard deviation. It is from this standard deviation that the confidence level (CL) (see equation 2) is derived.

The sample graph showing the errors between the prediction and the target and what was actually achieved through the implementation of genetic algorithm optimiser and the target is in Figure 7. The error shown is the sum-of-square errors of the normalised output and target.

The output are normalised between 0 and 1 and the standard deviation that is used to calculate CL is obtained from the normalised output. Each output gives its own standard deviation and the CL is calculated from the average standard deviations of all the members of the output vector. The mean square error (MSE) before the implementation of the optimisation method is 0.71 while the achieved MSE after optimisation is 0.41 (see Figure 7). The results from Figure 7 show that the proposed control framework is robust and improves the end of fermentation results.

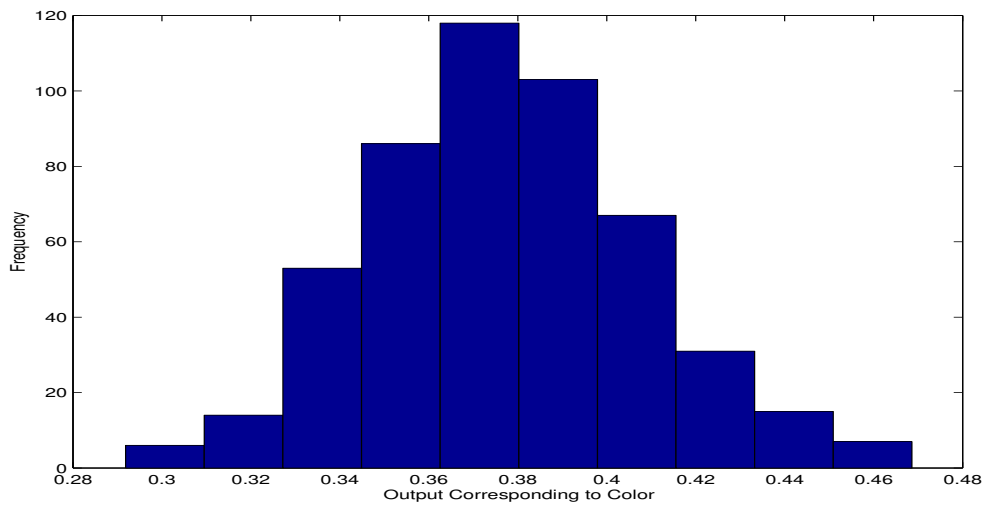

Fig. 6. Distribution of the output corresponding to colour level.

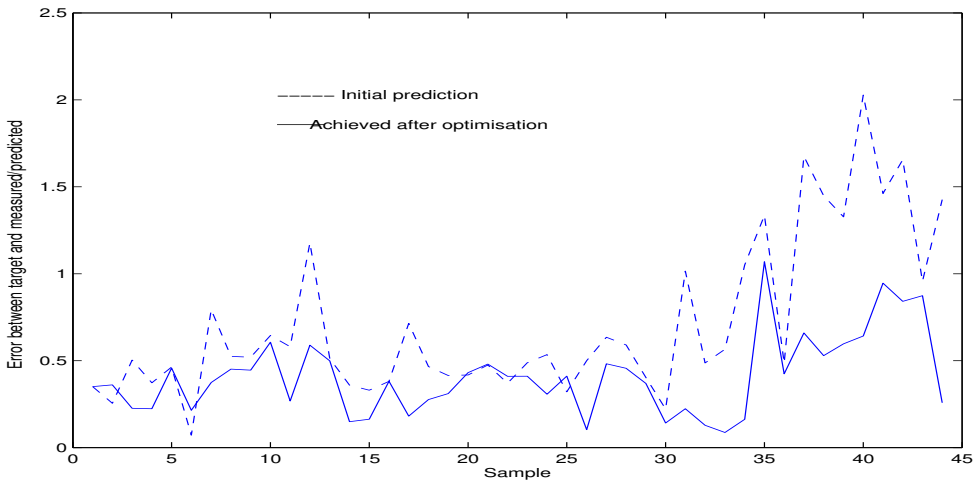

Fig. 7. A graph showing the error of the prediction from the initial inputs and the error achieved when the optimal inputs have been determined from the optimisation process

## 7. CONCLUSION

In this paper a control procedure that uses Bayesian neural networks and genetic algorithm is proposed to solve complex problems. The Bayesian networks are trained using the hybrid Monte Carlo method. The objective function used in the genetic algorithm optimiser minimises the error and maximises the confidence levels of the prediction. When the proposed procedure is implemented to control the fermentation process, it is observed that the procedure is able to give better results than was predicted without the use of genetic algorithm optimiser.

## REFERENCES

Bishop, C.M. (1995). *Neural Networks for Pattern Recognition*, Oxford University Press.

Blonbou, R, A. Laverdant, S. Zaleski and P. Kuentzmann. (2000) Active adaptive combustion control using neural networks. *Combustion Science and Technology*, **156**, 25-47.

Chang, S.S., W.H. Fu and F.S. Wen (1998). Load frequency control using genetic-algorithm based fuzzy gain scheduling of PI controllers. *Electronic Machines and Power Systems*, **26**, 39-52.

Fengyuan, R., R. Yong and S. Xiuming (2002). Design of a fuzzy controller for active queue management, *Computer Communications*, **25**, No. 1, pp. 874-883, 2002.

Goldberg, D.E. (1989). *Genetic algorithms in search, optimization and machine learning*, Addison-Wesley, Reading, MA.

Holland, J. (1975). *Adaptation in natural and artificial systems*, University of Michigan Press.

Johansson, B. and B. Hahn-Hägerdal (2002). The non-oxidative pentose phosphate pathway controls the fermentation rate of xylulose but not of xylose in Saccharomyces cerevisiae TMB3001, *FEMS Yeast Research*, **2**, 277-282.

Metropolis, N., A.W. Rosenbluth, M.N. Rosenbluth, A.H. Teller and E. Teller (1953). Equations of state calculations by fast computing machines. *Journal of Chemical Physics*, **21**, 1087-1092.

Michalewicz, Z., and D. Dasgupta, editors, (1997) *Evolutionary algorithms in engineering applications*, Springer-Verlag, New York.

Morimoto, T. and Y. Hashimoto (1996) Optimal control of plant growth in hydroponics using neural networks and genetic algorithms. *In W. Day, & P. C. Young (Eds.), Acta Horticulture Process, Mathematics and control application in agriculture and horticulture*. **406**, 433-440.

Neal, R.M. (1993). *Probabilistic inference using Markov chain Monte Carlo methods*, University of Toronto Technical Report CRG-TR-93-1, Toronto, Canada.

O'Connor, B., C. Riverol, P. Kelleher, N. Plant, R. Bevan, E. Hinchy, J. D'Arcy (2002). Integration of fuzzy logic based control procedures in brewing. *Food Control*, **13**, 23-31.

Öcten, G. (2002). Random sampling from low-discrepancy sequences: applications to option pricing. *Mathematical and Computer Modelling*, **35**, 1221-1234.

Pallaschke, D. (1997). *Foundations of Mathematical Optimization - Convex Analysis without Linearity*, Kluwer Academic Publishers, London.

Schurgel, K. (2001). Progress in monitoring, modelling and control of bioprocesses during the last 20 years, *Journal of Biotechnology*, **85**, 149-173.

Takaishi, T. (2002). Higher order hybrid Monte Carlo at finite temperature. *Physics Letters B*, **540**, 159-165.